\documentstyle[epsf]{mn}
\newcommand{\kms}{{\rm\ km\ s}$^{-1}$}

\newcommand{\pccm}{{\rm\ cm}$^{-3}$}
\newcommand{\oiii}{[O\,{\sc iii}] 5007-\AA\ }
\newcommand{\etal}{{et al.\,}}

\newcommand{\x}{{\times}}
\newcommand{\ha}{H$\alpha$}
\newcommand{\us}{_{\rm s}}
\newcommand{\K}{{\rm K}}
\newcommand{\uc}{_{\rm c}}
\newcommand{\uz}{_{\rm z}}
\newcommand{\mpr}{m_{\rm p}}

\def\ee#1{\times10^{#1}}

\title[Jet Cocoons and the Formation of Narrow Line Clouds in Seyfert Galaxies]
{
Jet Cocoons and the Formation of Narrow Line Clouds in Seyfert Galaxies}
\author[Steffen et al.]
       {
W. Steffen$^{1}$, J.L. G\'omez$^{1,2}$, R.J.R. Williams$^{1,4}$, A.C. Raga$^{3}$, A. Pedlar$^{5}$ \\
    $^{1}$Department of Physics and Astronomy, University of Manchester, Schuster Laboratory, Oxford Road, Manchester M13 9PL, UK \\
$^{2}$Instituto de Astrof\'{\i}sica de Andaluc\'{\i}a, CSIC, Apdo. 3004, Granada 18080, Spain \\
$^{3}$Instituto de Astronom\'{\i}a, UNAM, Apdo. Postal 70-264, 04510 Mexico, D.F., Mexico \\
$^{4}$Dept. of Physics and Astronomy, The University, Leeds LS2 9JT, UK \\
 $^{5}$Nuffield Radio Astronomy Laboratories, University of Manchester, Jodrell Bank, Macclesfield, Cheshire SK11 9DL, UK \\
}
\date{}

\begin{document}

\maketitle

 \begin{abstract}
 We present non-adiabatic hydrodynamic simulations of a supersonic
 light jet propagating into a fully ionized medium of uniform density
 on a scale representative of the narrow line region (NLR) in
 Seyfert galaxies with associated radio jets. In this regime the
 cooling distance of the swept up gas in the bowshock of the jet is of
 the same order as the transverse extent of the jet bowshock, as
 opposed to the more extreme regimes found for more powerful
 adiabatic large scale jets or the slow galactic jets which have been
 simulated previously. We calculate the emissivity for the \ha\ line
 and radio synchrotron emission. We find that the structure of the line
 emitting cold envelope of the jet cocoon is strongly dependent on the
 non-stationary dynamics of the jet head as it propagates through the
 ambient medium. 
 We observe the formation of cloud-like high density
 regions which we associate with NLR clouds and filaments.  We find
 that some of these clouds might be partially neutral and represent
 sites of jet induced star formation.  The calculated \ha\ flux and the
 spectral line width are consistent with NLR observations.  The
 simulation of the radio-optical emission with radiative cooling
 confirms the basic result of the geometric bowshock model developed
 by Taylor \etal\ (1989) that the start of noticeable optical line
 emission can be significantly offset from the hotspot of the radio
 emission.  However, the time-dependent nature of the jet dynamics
 implies significant differences from their geometric bowshock model.
 
\end{abstract}


\section{Introduction} 

The study of the Narrow Line Region (NLR) in Seyfert galaxies is
contributing substantially to the understanding of the central engine
of active galaxies. Of particular interest is the apparent close
association between radio continuum and optical line emission in this
region, which has a typical extent of 0.1-1\,kpc. The link between
optical and radio emission in the NLR manifests itself most
prominently in a positional association between linear radio
structures and optical line emission as well as in a correlation
between line width and radio power (Wilson and Willis 1980, Whittle
1985). It is widely accepted that the radio emission is related to the
outflow of relativistic plasma from the core of the galaxy. As is
found in higher power radio galaxies and quasars, this outflow appears
to be highly collimated into jets. The radio-optical association
suggests that the interaction of the jets with the interstellar medium
(ISM) strongly influences the dynamics of the ionized gas in the
NLR\@. It can be hoped that the study of this interaction will reveal
important information on the physical conditions of the gas in the NLR
and in the plasma jets.

The expansion of the cocoons of large scale extragalactic jets in
radio galaxies has been examined analytically by Scheuer (1974) and
Begelman and Cioffi (1989) and numerically by Cioffi and Blondin
(1992).  Blandford and K\"{o}nigl (1979) first suggested that the
interaction of jets with BLR clouds could cause the radio emission and
the associated kinematics of the BLR\@. The observation of symmetric
double or triple radio sources associated with emission line gas in
the NLR of a number of Seyfert galaxies (in particular in NGC\,5929,
Haniff, Wilson and Ward 1988, Su \etal 1996, and NGC\,1068, Wilson and
Ulvestad 1987, Gallimore \etal 1996) led to the development of more
detailed models invoking the interaction of plasmons or jets with the
local ISM\@. Pedlar \etal\ (1985) proposed an expanding plasmon model
for the NLR\@. They suggested that a high pressure relativistic plasma
bubble expands and drives a shock wave sweeping up the surrounding
ISM.

Whittle \etal\ (1986) found a spatial separation of
$0.95\pm0.11$\arcsec\ between the peaks of the \oiii components in
high dispersion spectroscopic observations. This separation is
significantly smaller than the measured distance between the radio
components. Based on these observations, Pedlar \etal (1987) proposed
a variant of the plasmon model similar to the radiative
bowshock model suggested earlier by Wilson \& Ulvestad (1987) for the
radio structure observed in NGC\,1068.  This model explains the
difference in the spatial separation between the peaks of the radio
and the optical components. It invokes a supersonic motion of a jet or
plasmon inducing a bowshock in the ambient medium. In the frame of the
head of the bowshock, the shocked ISM gas then moves along its
surface. All the basic properties of an expanding plasmon model are
still applicable, but, in addition, the cooling time of the gas behind
the bowshock allows the gas to flow a significant distance from the
vertex before cooling to a temperature sufficiently low ($\sim 10^4$
K) that optical lines are emitted. Since the radio emission is
expected to peak at the vertex, there will be a difference in the peak
positions of the optical and the radio emission. This model was
further developed by Taylor \etal (1989, 1992) and Ferruit \etal
(1996).

Recent Hubble Space Telescope (e.g. Capetti \etal 1995a,b; Bower \etal
1994, 1995) and MERLIN observations (Pedlar \etal 1993; Kukula \etal
1996) revealed overwhelming new structure in the NLR of several
Seyfert galaxies with radio jets. Discrete clouds and diffuse line
emission are found to be closely aligned with highly structured and
sometimes distorted radio jets.  These observations clearly show that
the idea of a radio jet interacting with the NLR can account for 
basic features of these jets. However, the phenomena are far more
complicated than can be accounted for by the simple geometric bowshock
model.  The full non-stationary character of the interaction between a
jet and its environment has to be taken into account to explain the
newly observed features.

In this paper, we therefore apply high resolution axisymmetric
hydrodynamic simulations to explore the time dependent interaction of
Seyfert jets with the NLR\@. We demonstrate that the association
between the radio and optical emission can be explained as a natural
consequence of the expansion of a hot jet cocoon into the interstellar
medium creating an envelope of dense cool gas and discrete emission
line knots which can be associated with the narrow-line clouds
themselves.  We compare our model with observations of the NLR
\ha\ emission in MKN~1066, which has a similar linear extension
(Bower \etal 1995).

In Section \ref{analytic.sec} we estimate the properties of our cocoon
shock model analytically.  The numerical procedure of the simulations
is briefly described in Section
\ref{simulation.sec}.  Section \ref{results.sec} contains the results 
and their discussion. We summarize our conclusions in Section
\ref{conclusions.sec}.

\section{Analytical estimates}
\label{analytic.sec}

We consider the dense optically emitting envelope of swept-up ISM gas
around the cocoon of a supersonic jet propagating into a uniform,
ionized medium. The ionization is assumed to be due to photoionization
from a UV source at the active nucleus of the galaxy. For a
UV photon rate $S\sim 10^{52}{\rm s^{-1}}$ similar to the one in NGC5929
(Bower \etal 1995) and an ambient density of 1\,cm$^{-1}$ the radius of
the Str\"omgren sphere is 700\,pc, which is well beyond the distances
we consider in this paper. For an estimate of the initial conditions
in a simulation which can be compared with typical observations, we
assume that the envelope is a thin cylindrical shell of radius $r\uc
\sim$ 20\,pc, length $l\uc$ = 150\,pc,
\ha\ luminosity $L_{\rm H\alpha} = 10^{37}-10^{39}$erg\,s$^{-1}$, and
a thickness $d\uc \ll r\uc$. Within our model the spectral line width
in the NLR will be similar to twice the cocoon expansion speed, if
observed from a direction perpendicular to the jet axis. We
therefore have typical shock velocities of 100--200\,\kms
(e.g. MKN\,1066, Bower \etal 1995).  In the following, we estimate
some characteristic quantities, such as the density $n_0$ of the ambient
ISM, the density $n\uc$ and thickness $d\uc$ of the dense cool
envelope, using this range of observed sizes and luminosities.

\subsection{Non-magnetic case}
\label{non-magnetic.sec}

We first assume that there is no magnetic field in the environment
which could limit the compression at the shock, which is driven
perpendicularly to the jet axis into the ISM by the overpressured
cocoon. We further assume that the shock is isothermal, at the
temperature $T_0 = 10^4$K of the pre-shock gas. This assumption
is justified by estimating the cooling distance $\delta$
behind a shock of speed $v\us = $ 150\kms\ in typical environmental gas
with a hydrogen density of $n_0 = $1\pccm, using the adiabatic
compression factor of 4. We then have (Taylor, Dyson and Axon 1992)
\begin{equation}
\label{dcool.eq}
\delta = \frac{\bar{m} v\us k T\us^{(1-\alpha)}}{4 \mpr n_0 \Lambda_0},
\end{equation}
where we use a power-law cooling function of the form
\begin{equation}
\label{fcool.eq}
\Lambda (T\us) = \Lambda_0 T\us^\alpha
\end{equation}
with $\Lambda_0 = 4.6 \cdot 10^{-18} {\rm erg\,s}^{-1}{\rm cm^3}$ and
$\alpha=-0.76$ at $T > 1.5 \cdot 10^5$K.  The 
temperature $T_s$ immediately behind the shock is given by
\begin{equation}
\label{shocktemp.eq}
T\us = \frac{3}{16} \frac{\bar{m}}{k} v\us^2,
\end{equation}
where $\bar{m}$ is the mean molecular weight and $k$ is the Boltzmann
constant (we assume $\bar{m} =0.5 \mpr$, $\mpr$ is the proton mass).
The cooling distance perpendicular to the shock is then $\delta =
1.8\x 10^{17}$~cm or 0.06~pc with corresponding postshock temperature
of $T\us = 2.5\x 10^5$~K. For $v\us$=100 (200)\,\kms\ the cooling
distance will be 2.2 (3.7) times smaller (larger) than for
$v\us$=150\,\kms. All these values are much smaller than the radius
$r\uc\sim 20$pc of the cocoon, justifying the assumption of an
isothermal shock for the estimates.

We now calculate the density $n\uc$ and thickness $d\uc$ of the compressed
postshock layer of gas and the density of the environmental gas $n_0$. We 
start with the compression ratio of the isothermal shock, which is
\begin{equation}
\label{f1.eq}
f = \frac{n\uc}{n_0}
  = \frac{\bar{m} v\us^2}{kT_0}
  = 120 \, \frac{\bar{m}}{\mpr}
        \left(\frac{T_0}{10^4\K}\right)^{-1}
        \left(\frac{v\us}{100\,{\rm km\,s^{-1}}}\right)^2
\end{equation}
In terms of the volumes of the cylindrical cocoon
$V\uz$ and the cylindrical shell of swept up gas around the cocoon
$V\uc$ we also have
\begin{equation}
\label{f2.eq}
f = \frac{V\uz}{V\uc} 
  = \frac{\pi r\uc^2 l\uc}{2\pi r\uc d\uc l\uc}
  = \frac{r\uc}{2 d\uc}.
\end{equation}
Combining equations (\ref{f1.eq}) and (\ref{f2.eq}) we find the 
thickness of the layer of swept up gas to be 
\begin{eqnarray}
\label{dc.eq}
d\uc &=& \frac{r\uc}{2} \frac{kT_0}{\bar{m} v\us^2} \nonumber \\
    &=& 0.083 \,{\rm pc} \,\frac{\bar{m}}{\mpr}
                     \left(\frac{r\uc}{{20 \rm pc}}\right)
                     \left(\frac{T_0}{10^4\K}\right)
        \left(\frac{v\us}{100\,{\rm km s^{-1}}}\right)^{-2}
\end{eqnarray}
Assuming pure hydrogen gas of uniform density, the \ha\ luminosity of this
layer will be
\begin{eqnarray}
L_{\rm H\alpha} &=& n\uc^2 \epsilon V\uc \nonumber \\
  &=& 3.7\x10^{38} {\rm erg\,s^{-1}} 
      \left(\frac{n\uc}{100 {\rm cm^{-3}}}\right)^2
      \left(\frac{V\uc}{10^{59} {\rm cm^3}}\right)
\end{eqnarray}
where $\epsilon=3.8\x 10^{-25}{\rm erg cm^3 s^{-1}}$ is the \ha\
emission coefficient at $T_0$ (see Section \ref{optical.sec}).

The total mass swept up from the environment into the shell is
\begin{equation}
\label{M.eq}
M = n\uc \mpr V\uc = 8300 {\rm M_\odot} 
        \left(\frac{n\uc}{100 {\rm cm^{-3}}}\right)
        \left(\frac{V\uc}{10^{59} {\rm cm^3}}\right)
\end{equation}
We assume full ionization of the gas in the NLR\@. A column of swept up 
hydrogen gas of density $n\uc$ at a distance $l\uc$ from the ionizing photon
source with a photon flux $S$ will stay fully ionized up to a thickness \\
\begin{equation}
d\us = 1.1{\rm pc}
        \left(\frac{n\uc}{100 {\rm cm^{-3}}}\right)^{-2}
        \left(\frac{l\uc}{100 {\rm pc}}\right)^{-2}
        \left(\frac{S}{10^{52} {\rm s^{-1}}}\right)
\label{stroem.eq}
\end{equation}
assuming that no photons are intercepted on their way to the cloud.

In Section \ref{results.sec} we compare
these analytic estimates with the results from our simulation.

\subsection{Magnetic case}
\label{magnetic.sec}

We now consider the case where the undisturbed environment
is threaded by a random magnetic field, assuming 
the components in the plane of the shock will be compressed
with the field lines frozen into the gas. The effect of the magnetic
field will be to limit the compression of the shocked gas when the magnetic
pressure equals the thermal pressure. The compression
factor $f_m$ is then given by (e.g. Dopita and Sutherland 1995):
\begin{eqnarray}
f_m &=& \frac{n_{cm}}{n_{0m}} 
     =  \frac{(8\pi\bar{m})^{1/2} v_s}{B_{\parallel} n_{0m}^{-1/2}} \nonumber \\
    &=& 64 \frac{\bar{m}}{\mpr}
           \left(\frac{v\us}{100\,{\rm km\,s^{-1}}}\right)
           \left(\frac{B_{\parallel} n_{0m}^{-1/2}}{{\rm \mu G\,cm^{-3/2}}}\right)
\label{fm.eq}
\end{eqnarray}
The thickness of the shocked gas layer is then
\begin{eqnarray}
d_{\rm cm} &=& \frac{r\uc}{2 v\uc}\frac{B_{\parallel} 
               n_{\rm 0m}^{-1/2}}{(8\pi\bar{m})^{1/2}} \nonumber \\
    &=& 0.15\,{\rm pc} \,\, \bar{m} \mpr^{-3/2}
        \left(\frac{r\uc}{{20 \rm pc}}\right)
        \left(\frac{v\us}{100\,{\rm km\,s^{-1}}}\right)^{-1} \nonumber\\
    &&\left(\frac{B_{\parallel} 
     n_{{\rm 0m}}^{-1/2}}{{\rm \mu G\,cm^{-3/2}}}\right)
.
\label{dcm.eq}
\end{eqnarray}
Values of the magnetic parameter
$(B_{\parallel} n_{{\rm 0m}}^{-1/2})/({\rm \mu G\, cm^{-3/2}})$ are of order 1,
in the range 1-10 (Dopita \& Sutherland, 1995).

Due to the additional magnetic pressure, the thickness of the cold
envelope will be larger and the density lower than in the non-magnetic
case. Since the optical line emission depends on the square of the
density, the emission received from a strongly magnetized envelope
will be considerably lower than  that coming from a non-magnetic envelope.

\section{Numerical simulation}
\label{simulation.sec}

\subsection{Initial conditions}
\label{initcond.sec}

The determination of jet parameters like plasma speed and density is a
classical and basically unsolved problem in extragalactic
astrophysics. For the purpose of simulating the the expanding
envelope, the exact values of the jet speed and density input
parameters are not important.  The expansion of the cocoon shock is
principally determined by the cocoon pressure, and a given pressure
can be obtained from a range of jet parameters. For practical purposes
this range of parameters is also somewhat restricted by the available
computing resources. We therefore performed a small series of test
runs varying the particle density in the ISM around 1\pccm, jet
velocities around 10$^4$\,\kms (Bicknell \etal 1990) and Mach
numbers around 10. A Mach number higher than 5 is required for a jet
with a noticeable cocoon (Norman, Smarr and Winkler 1985). From these
tests we chose a set of parameters which compares well with the
observations of Seyfert galaxies given in Section
\ref{analytic.sec} (in particular MKN\,1066) and which illustrates
best the qualitative features of the formation and structure of the
dense cocoon envelope. A more detailed account on the influence of
varying the parameters will be given in a forthcoming paper (Steffen
\etal, in preparation). The chosen parameters are:

\begin{eqnarray}
n_0 &=& 1.5 {\rm\ cm}^{-3} \nonumber\\
T_0 &=& 10^4 {\rm K} \nonumber\\
n_j &=& 0.18 {\rm\ cm}^{-3} \nonumber\\
M_j &=& 8.3 \nonumber\\
v_j &=& 6.8\x 10^8 {\rm\ cm~s}^{-1} \nonumber\\
r_j &=& 5.3\x 10^{18} {\rm cm} \nonumber
\end{eqnarray}

This yields a jet mechanical luminosity of $L_j = 2\times 10^{39}
{\rm erg\,s^{-1}}$.

\subsection{Scaling and similarity}
\label{scaling}

In the absence of cooling, the jet will be evolve in a self-similar
fashion 
\begin{equation}
l_c = {\cal A}\left(L_j t^3\over\bar{m} n_0\right)^{1/5}
\label{e:self}
\end{equation}
with ${\cal A} \simeq 1$, once $l_c \gg (n_j/n_e)^{1/2} r_j =
2\ee{18}{\rm\,cm}$ (Falle 1991).

Cooling will begin to have an effect when the advance speed of the
outer shock at the base of the jet is roughly $100{\rm\,km\,s^{-1}}$
(Dyson 1984), that is roughly when $l_c = l_{\rm cool} =
6\ee{19}{\rm\,cm}$, if the aspect ratio of the jet is ${\cal R} =
l_c/r_c = 5$.  Thus, for our typical parameters, the jet will be self
similar before cooling becomes important (neglecting the effects of
turbulence in the jet).

Once cooling becomes non-negligible, self-similarity no longer holds
but the evolution of jets with different parameters will be similar to
each other, if times are scaled to the time at which the gas begins to
cool, and distances are scaled to the length of the jet at this time.
For the cooling law, equation (2), the lengths will scale as
\begin{eqnarray}
l_{\rm cool} &\simeq& 6\ee{19}{\rm\,cm}
\left(L_j\over2\ee{39}{\rm\,erg\,s^{-1}}\right)^{(3-2\alpha)/(9-4\alpha)} 
\nonumber \\
&&\cdot \left(n_0\over1.5{\rm\,cm^{-3}}\right)^{-(6-2\alpha)/(9-4\alpha)}
\end{eqnarray}
and the times as
\begin{eqnarray}
t_{\rm cool} &\simeq& 6\ee{11}{\rm\,s}
\left(L_j\over2\ee{39}{\rm\,erg\,s^{-1}}\right)^{(2-2\alpha)/(9-4\alpha)} 
\nonumber \\
&&\cdot \left(n_0\over1.5{\rm\,cm^{-3}}\right)^{-(7-2\alpha)/(9-4\alpha)}.
\end{eqnarray}
If the jet power is doubled, the timescales will increase by 22 per
cent and the lengthscales by 30 per cent, while if the ambient density
is doubled the lengthscales will decrease by 35 percent and the
timescales by nearly 40 per cent.  

Once the cooling becomes rapid at the head of the jet, when $l_c
\simeq {\cal R}^{3/2} l_{\rm cool}$, the global evolution of the jet
will enter a second self-similar regime with its evolution varying
according to equation~(\ref{e:self}), only with a rather smaller
constant of proportionality.

\subsection{The hydrodynamic code}
\label{code}

We used the adaptive grid hydrodynamic code described by Biro \etal
(1995) in axisymmetric mode.  It solves the equations of mass,
momentum and energy conservation using a flux-vector-splitting
scheme. The computation was carried out in a 5-level, binary adaptive
grid. In our simulations we assume full ionization of the jet and
ambient medium due to photoionization from the central
UV source. We use the non-equilibrium cooling function described by
Biro \etal (1995) with an additional term taking into account
bremsstrahlung losses, $L_b$, given by
\begin{equation}
\label{bems.eq}
L_b = 2.29\x10^{-27} n_e n_p T_e^{\frac{1}{2}}\ {\rm erg\,s^{-1}~cm^{-3}}
\end{equation} 
where $n_e$ and $n_p$ are the electron and proton densities in
particles per cm$^{3}$, respectively, and $T_e$ is the electron
temperature in Kelvin (Cox \& Tucker 1969).  
We do not allow the gas temperature to drop below $10^4$~K, where we
assume it is maintained by the thermostat effect of \oiii emission.

The computational domain was set to be 513x1025 computational cells
and $5\x 1.25$ times $10^{20}$cm. The grid cells are smaller in radial
direction than in axial direction for better resolution of the small
distance scales of the radially expanding envelope. The initial jet
radius was covered by 27 cells in the axisymmetric simulation. The
cooling distance $\delta$ was marginally resolved for $v_s = 150$\kms.
The thickness $d\uc$ of the cold envelope of swept up gas was near the
resolution limit and was smeared out artificially over 2-3 cells at
shock speeds higher than 150\kms.  This artificially limits the
compression with effects similar to those expected if magnetic fields
in the ISM were present (with a magnetic parameter of the order of 5,
although magnetic fields were not taken into account
explicitly). However, increasing the resolution further would lead to
prohibitively large computing times for these initial
studies. Quantitative results will therefore be only order of
magnitude estimates, but the qualitative results, as discussed in this
paper, will not be significantly altered. The boundary conditions are
reflective on the axis and on the left side of the computational
domain (except for the inflow condition where the jet is injected, see
Fig.\,\ref{dp.fig}). The top and right boundaries have outflow
conditions.

\subsection{Emission maps}
\label{emission.sec}

\subsubsection{Optical emission}
\label{optical.sec}

We calculate the \ha\ emissivity $\epsilon$ using radiative
recombination (Case B) following Aller (1984).  In the temperature
regime of our simulations ($T>10^4$K) only recombination contributes
considerably to the emissivity. 
We assume full ionization of the
hydrogen, which would mainly be due to photoionization from the UV
source at the centre of the galaxy and, to some extend, from shock
ionization in the bowshock region.
\begin{equation}
\epsilon = 4.16\times 10^{-25} {\rm erg\,cm^3\,s^{-1}} 
           (T_4^{0.983} 10^{0.0424/T_4})^{-1} 
\end{equation}
where $T_4$ is the temperature in units of $10^4$K.

\subsubsection{Radio emission}
\label{radio.sec}

  Only relativistic electrons (and perhaps positrons) contribute
significantly to the radio emission of jets. Since it is still unclear
how these high energy electrons are generated (most probably through
re-acceleration in shocks), we assume that this population of fast
electrons shares the same hydrodynamics as the non-relativistic fluid
of electrons and protons. A more detailed description would require
the consideration of two jet populations to account for the
non-relativistic gas (electrons and protons) with ratio of specific
heats of 5/3, and the ultrarelativistic electrons with ratio of 4/3.
Duncan, Hughes, Opperman (1996) have performed preliminary simulations
of highly relativisitic jets using a variable adiabatic index and
found that the rest frame variables vary smoother along the jet axis,
with smaller amplitude maxima at the bowshock, and in the vecinity of
the contact surface. We ignore these effects in the present simulations.

  In order to calculate the synchrotron emission from the jet whose
hydrodynamics is modeled as above, we need to establish how the internal
energy is distributed among the relativistic electrons. Observations suggest
that this distribution follows the usual power law $N(E) \, {\rm d}E = N_o
E^{-p} {\rm d}E$, with spectral index $p$, and energies above a minimum value
$E_{\rm min}$. Assuming that the relativistic electron energy density ${\cal
U}$, and number density ${\cal N}$ are proportional to their corresponding
thermal quantities calculated by the hydrodynamical code as a function of
position in the jet, the previous power law is determined by the equations
(G\'omez et al. 1995)
\begin{equation}
\label{ncero}
N_o = \left[{\cal U} \: 
(p\!-\!2)\right]^{p-1}
\left[{\cal N}\:(p\!-\!1)\right]^{p-2}
\end{equation}
and
\begin{equation}
\label{emin}
E_{\rm min} = \frac{\cal U}{\cal N}\:\:
\frac{p\!-\!2}{p\!-\!1} ,
\end{equation}

The synchrotron emission is also determined by the magnetic field, and since
we are neglecting its influence on the fluid dynamics, we assume that the
magnetic energy density remains a fixed fraction of the particle energy
density, which leads to a field of magnitude proportional to ${\cal U}^{1/2}$.
We refer the reader to G\'omez et al. (1993, 1995) for a detailed
discussion of the radio emission calculations.

\section{Results and Discussion}
\label{results.sec}

\subsection{Structure}
\label{structure.sec}

We show the results of our simulation at a time when the jet has
reached a distance of approximately 150~pc from the injection point.
This is 127\,000 years after the jet switched on.

\begin{figure*}
\centering
\mbox{\epsfxsize=5.6in\epsfbox[60 200 495 639]{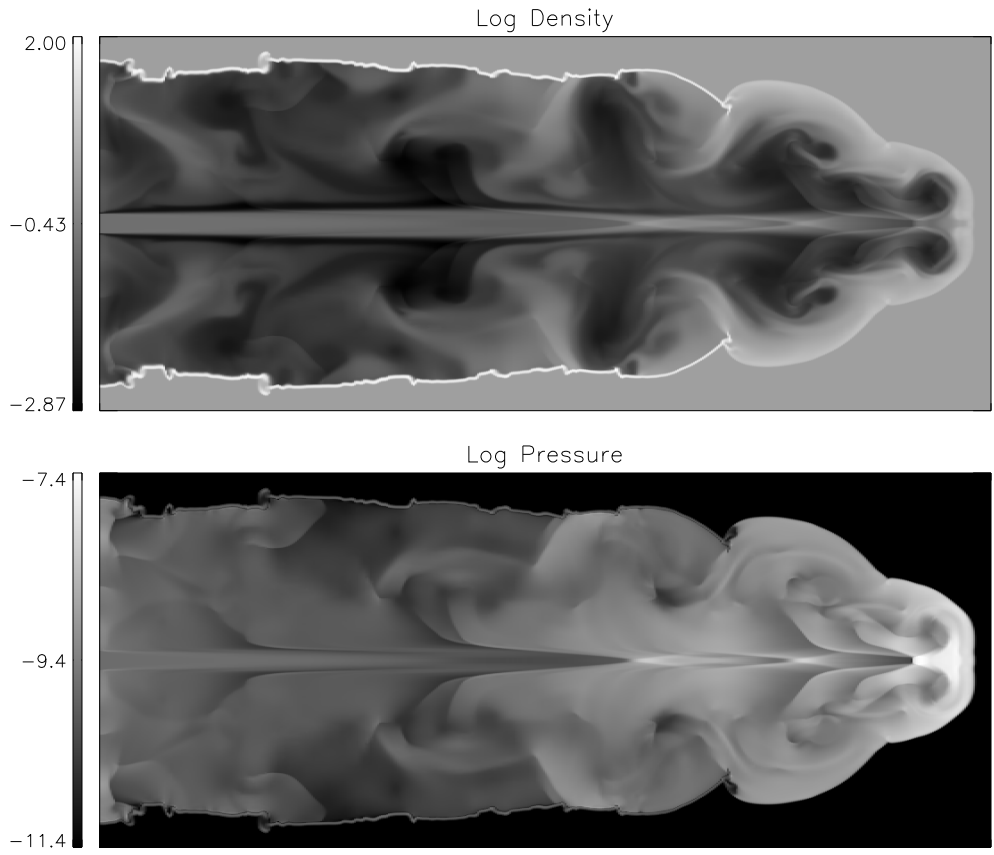}}
\caption{Density (top panel) and pressure (bottom panel) distribution on 
a logarithmic scale after 127~kyr. Note the high-density cocoon
envelope after the swept up ISM-material has cooled to the equilibrium
temperature of 10$^4$~K. The recollimation shocks and the very high pressure
near the jet shock can clearly be seen in the pressure image.}
\label{dp.fig}
\end{figure*}

\begin{figure*}
\centering
\mbox{\epsfxsize=5.6in\epsfbox[0 0 557 399]{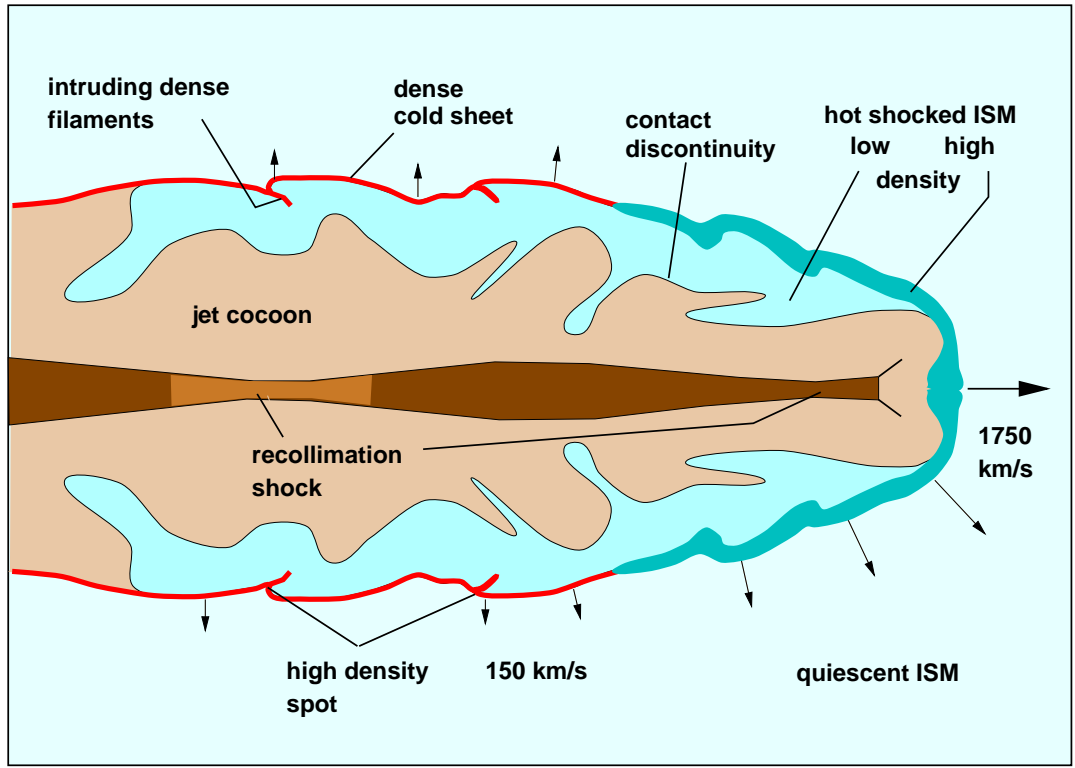}}
\caption{A schematic summary of the features seen in our hydrodynamic
simulations. It includes the jet with recollimation shocks, the
jet-shock, the cocoon of jet plasma, the hot bowshock and the cold layer
shocked ISM.}
\label{jetscheme.fig}
\end{figure*}

Fig.\,\ref{dp.fig} shows the density (top) and the pressure (bottom)
distribution on a logarithmic scale.  We can identify several
characteristic features which are marked in the schematic view shown
in Fig.\,\ref{jetscheme.fig}. The jet itself shows several
recollimation shocks on the axis, best seen in the pressure image. At
the tip of the jet, the termination shock generates high pressure
and high temperature jet plasma. This plasma flows back into the hot
jet cocoon, which surrounds the high speed jet. The jet cocoon is
surrounded by a thick layer of hot gas from the
interstellar medium which has passed through the bowshock. Near the head
of the jet we find a thin layer of hot shocked gas from the ISM which
further back turns into an even thinner, but cold and very dense
envelope of the jet cocoon.  This envelope is what distinguishes the
NLR jet from adiabatic high power extragalactic jets.
The fact that this envelope starts at a significant distance from the front
of the bowshock distinguishes NLR jets from the regime of strongly 
cooling galactic jets (e.g. Blondin, Fryxell and K\"onigl 1990).

We find that the radius of the cold envelope in the simulation is
approximately $7.8\times 10^{19}{\rm cm}$ and the expansion speed of
the cocoon is somewhat above 100~\kms. The peak density in the
envelope varies from values between 30 to 40 cm$^{-3}$ over most of
its extent, up to 100 cm$^{-3}$ in the dense clouds. This
corresponds to compression factor of 30 and 70, respectively. This is
factors of 2 too low compared to Eq.\,(\ref{f1.eq}), which is
probably due to numerical smearing. This effect
mimics a magnetic parameter of the order of 5 (see Eq.\,\ref{dcm.eq}),
which corresponds to a ambient magnetic field near $5\mu G$.  The
integrated \ha\ luminosity at this timestep is $5.8\x 10^{37} {\rm erg\,
s^{-1}}$ and the envelope has a thickness of $d\uc \sim 1-2\x10^{18}$\,cm
and volume of $V\uc \sim 2\x 10^{59} {\rm cm^3}$.  These values are
in good agreement with our analytical estimates and the results from
optical observations of Seyfert galaxies with associated NLR radio
jets.

\begin{figure*}
\centering
\mbox{\epsfxsize=3.3in\epsfbox[0 0 259 372]{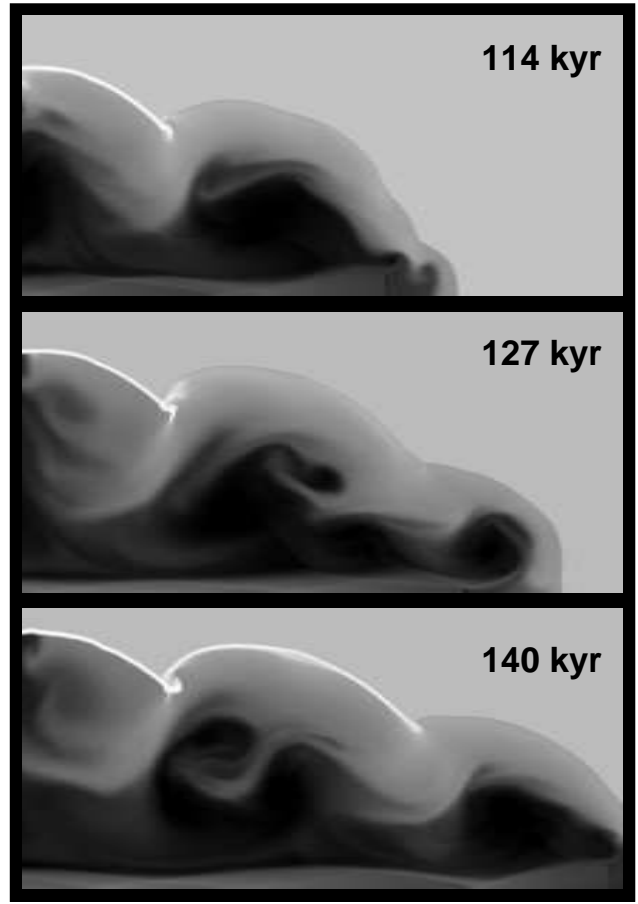}}
\caption{The bowshock structure (using the \ha\ emissivity) is shown at three
different times with the same separation between them. While the head
of the jet has advanced considerably between the first and second
panel, the point where the cold envelope starts has not advanced
appreciably. After a further timestep, a full arc has collapsed and
cooled and has become very bright. }
\label{collapse.fig}
\end{figure*}

We find that the non-stationary dynamics of jet propagation
through interstellar medium have a strong influence on the structure
of the cold shell. The quasi-periodic recollimations of the jet cause
the advance speed of the jet head to oscillate, thereby accumulating
jet material in a restricted nose region for some period of time. It
is then released into the cocoon almost explosively, when the jet
expands and slows down again.  This imposes a characteristic arc
structure onto the bowshock and the cold shell of swept up gas. These
arcs cool almost as a unity once catastrophic cooling sets in. This
can be seen in Fig.\ref{collapse.fig}, where the density is shown at
three different times with the same separation between them. While the
head of the jet has advanced considerably between the first and second
panel, the point where the cold dense envelope starts has not advanced
appreciably. However, later, after the same timestep, the full arc has
cooled. Therefore, a cooling distance for the bowshock measured
from the head of the jet, as  discussed by Taylor \etal (1989, 1992),
can at best be only an average property.

At the positions where neighbouring arcs meet, the two shock waves
cross at an angle near 90$^\circ$. This interaction will produce a Mach
shock at the vertex with an effective speed of
$\sqrt{2}$ times the speeds of the incident shocks (supposing that
they are the same). Correspondingly, the compression and the
emissivity of the gas increase by factors of $\sim 2$ and $\sim 4$,
respectively.  As a result of the radiative nature of the crossing
shocks they merge to form short high density filaments and clouds,
marked as ``high density spots'' and ``intruding dense filaments'' in
Fig.\,\ref{jetscheme.fig}. In the simulation the spots have peak
densities between 80 and 100\pccm, consistent with the previously
estimated additional compression by factors around 2. We suggest that
these spots could be identified with the brightest discrete clouds in
the NLR of some Seyfert galaxies. Note that the spacing between these
spots is directly related to the time dependent quasi-periodic
recollimation of the jet and therefore contains information about the
history of the interaction between the jet and its
environment. Several Seyfert galaxies show a knotty NLR structure and
the distances between the strongest emission line clouds is indeed
similar but rather smaller than the distances between the radio knots
(e.g. MKN~3, Capetti \etal 1995a; MKN~6, Capetti \etal 1995b;
NGC~4151, Boksenberg \etal 1995).

At some distance from the jet head the arc structure in the
envelope is lost and it becomes roughly cylindrical justifying the
approximations of Section \ref{analytic.sec}
(Fig.\ref{dp.fig}). However, the high density clouds and filaments
created at the intersection points of these arcs are retained. When
the envelope fragments (as can be expected in a full three dimensional
simulation) these regions will be seen as discrete NLR clouds and
filaments. Since the creation of these arcs and clouds is related to
the recollimations of the jet, we have found a direct relation between
the optical structure of the NLR and the internal non-stationary shock
structure of the jet.

The expansion of the overpressured jet cocoon into the undisturbed
interstellar medium is similar to a supernova explosion. Some
theoretical aspects can therefore be treated in an analogous
fashion. This applies in particular to the stability of the shock and
the postshock layer of swept up gas. The stability of radiative shock
wave has been considered by a number authors with varying emphasis
(e.g. Vishniac 1983, Bertschinger 1986).  The radiative shock
instabilities discussed by Bertschinger (1986) are present in our
simulations and produce the pressure variations near the cold
envelope. These are likely to be disruptive and thus contribute to the
formation of NLR clouds and filaments. From the results presented by
Jun, Norman and Stone (1995) it can be expected that the shock region
in our simulation is also Rayleigh-Taylor (R-T) unstable, with fastest
growth of features on the scale of the thickness of the cold layer of
swept up gas.The onset of these instabilities can be observed as
pressure variations and small ripples in some sections of the cold
envelope (Fig.\ref{dp.fig}). As these instabilities grow, they might
contribute to the fragmentation process of the envelope and form
further NLR clouds, in addition to the high density clouds produce at
the intersection of the arcs discussed in the previous
section. However, magnetic fields have a stabilizing effect and
therefore different combinations of ambient magnetic parameters and
jet powers might lead to two categories of evolved cocoon envelopes:
R-T stable narrow line regions with fewer small clouds and R-T
unstable ones with a high number of small NLR clouds with separations
of a few times the thickness of the cold layer (i.e.  $\sim$
1\,pc). More detailed investigations of this process will have to be
carried out to quantify this prediction over a range of parameters in
order to compare it with recent and future high resolution
observations of the NLR in Seyfert galaxies.

The Str\"omgren column given by Eq.(\ref{stroem.eq}) is similar to the
thickness of the cocoon envelope and the size of the clouds. It may 
well be that the highest density clouds can develope neutral cloud cores. 
As a result, NLR clouds could be sites of jet induced star formation
(Van Breugel and Dey 1993, Tresch-Fienberg \etal\ 1987).

\subsection{Emission}
\label{emission_res.sec}

\begin{figure*}
\centering
\mbox{\epsfxsize=5.6in\epsfbox[0 0 640 378]{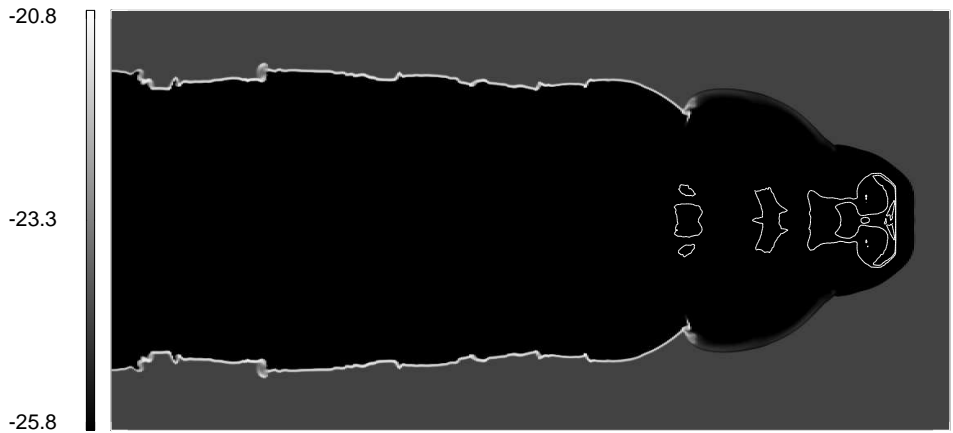}}
\caption{The \ha\ emissivity map is shown on a logarithmic scale
along with the calculated radio emission at 10\,GHz superimposed as
contours (radio contours are drawn for 0.6, 5 and 50 percent
of peak brightness).  Most of the optical emission comes from the thin
envelope of swept up and cooled ISM gas. Particularly high emissivity
regions are found where the bowshock `arcs' intersected when still
near the head of the jet. The radio emission is concentrated near the
head of the jet, where temperature and pressure of the jet plasma is
highest after passing through the jet shock.}
\label{ha.fig}
\end{figure*}

In Fig.\ref{ha.fig} the \ha\ emissivity distribution is displayed on a
logarithmic along with the radio
emission superimposed as contours in the lower panel.  As expected,
most of the optical line emission comes from the thin envelope of cold
postshock material. The \ha\ luminosity obtained from this
axisymmetric simulation is $5.8\times10^{37} {\rm erg\,s^{-1}}$. This
is well within our expectations considering the simplicity of our
analytical estimates. The length of the NLR jet in MKN~1066 is similar
to that in our simulation, although the transverse extent 
could be 2-3 times smaller. From observations by Bower \etal
(1995) we determined the \ha\ flux coming from the area covered by the
NE jet in MKN~1066 to be $3.3\times10^{37} {\rm erg\, s^{-1}}$, which
is in excellent agreement with our simulation, taking into account the
the simplifications of the model and the smaller radius of the cocoon
in MKN~1066.

>From our model it can be expected that higher power radio jets yield
faster cocoon shocks, producing broader emission lines. This will
result in a correlation between line width and radio power, which is
consistent with the observed correlation between these quantities
(Wilson and Willis 1980, Whittle 1985). Our initial study does,
however, not allow a more quantitative conclusion about the
correlation. An extensive survey of the parameter space and other
factors which influence the line width, like galaxy rotation, is
needed for a detailed theoretical study of this correlation in terms
of our model.

\begin{figure*}
\centering
\mbox{\epsfxsize=3.3in\epsfbox[40 300 580 780]{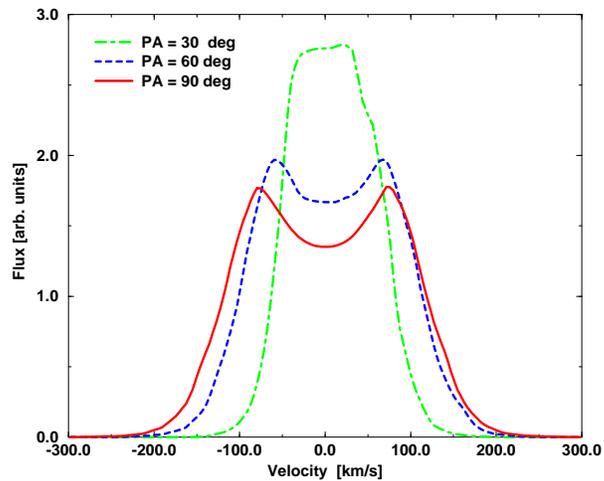}}
\caption{The spectral line shape of the \ha\ emissivity distribution 
as seen from different angles between the line of sight and the jet axis 
is shown. For large angles the line is double peaked, but for angles near 
the jet axis the shape is rather rectangular. }
\label{spec.fig}
\end{figure*}

The calculated line spectrum is shown in Fig.\ref{spec.fig} for three
different viewing angles. In the spectrum, emission from the
background has not been included in order to focus on the expanding
envelope. Inclusion of the background would produce a narrow peak at zero
velocity.  Within the volume of this simulation the background
emission amounts to $4.8\times 10^{36} {\rm erg\,s^{-1}}$, an order of
magnitude less than contained in the envelope. The strength of this
central peak, compared to the broader contribution from the
envelope, is a measure of the ratio of disturbed and
undisturbed gas.  Because of the cylindrical shape of the expanding
envelope and the small velocity of the gas along the shell, the line
shape of the envelope is double peaked for viewing angles far from the
axis but becomes almost rectangular if the line of sight is near the
jet axis. The line width of the expanding cylindrical shell is, of
course, not only a function of the expansion speed (which is
approximately the same as the shock speed), but depends also on the
angle between the symmetry axis and the line of sight. This is in
apparent disagreement with results from Whittle (1985), which suggest
that there is no strong correlation between the orientation of the
galaxy and the NLR line width, assuming that the jet follows the
rotation axis of the galaxy.  However, observations of radiation cones
(e.g. Wilson and Tsvetanov 1994), which at least can be expected to
have a closer link to the radio jets, and point symmetric bending of
jets in Seyferts (presumably as a result of the interaction with the
rotating environment, Wilson and Ulvestad 1982, Steffen \etal\ 1996)
suggest that the jets often are not aligned with the rotational axis
of the galaxies. This would strongly weaken any correlation between
line width and galaxy rotation.

The contours in Fig.\ref{ha.fig} show the radio emission at 10~GHz as
calculated from our model described in Section \ref{radio.sec}. A
spectral index in the power law energy distribution of the electrons
of 2.4, and magnetic field strength at the jet inlet of 100$\mu{\rm
G}$ have been used. A total flux for the jet of 4 mJy is obtained. As
expected, most of the radio emission comes from the hotspot region
near the head of the jet, with some low brightness emission from the
recollimation shocks. There are a number of Seyfert jets in which the
the head of the jet is dominant at radio wavelength. However, in most
cases there are a number of strong radio knots along the jet. More so
in the case of NGC~1066, which has rather continuous emission along
the jet (at the resolution of the VLA) and no hot spot near the
head. This highlights the problem of the nature of the knots seen
along radio jets in general. Internal shocks can produce this kind of
knots. In our simulations we find two different types of internal
shocks: the recollimation shocks and the shocks produced by the
interaction with the turbulent cocoon near the head of the jet. From
our simulations and our model for the synchrotron emission we find
that these are not strong enough to explain the radio emission knots
seen in most of the Seyfert jets. One possible solution are time
variations of the ejection parameters at the base of the jet. G\'omez
et al (1996), Hughes, Duncan, and Mioduszewski (1996) and Komissarov
and Falle (1996) have shown that variations of the ejection velocity
can produce strong internal working surfaces which visible as strong
radio knots in the jet. Such variations in the jet properties possibly
have some influence on the structure of the cold cocoon envelope and thus
on the NLR. In a forthcoming paper we will investigate the effects of 
variable jet properties on the NLR in more detail 
(Steffen et al, in preparation). 

\section{Conclusions}
\label{conclusions.sec}

Using hydrodynamic simulations with radiative cooling we confirm the
basic structure of the interaction between a jet and the NLR in
Seyfert galaxies as obtained in a geometric model by Taylor \etal
(1989).  However, we find new features of the interaction, like the
formation of NLR clouds and a larger extent of the cold cocoon
envelope.  Due to the time-dependent propagation speed of the jet into
the ISM, the cooling distance of the bowshock material as measured
from the head the jet varies episodically with time.  Our simulations show
that there might be a link between the time-dependent internal
recollimation structure of the radio jets and the narrow line
clouds. We expect that the distances between particularly bright knots
along the symmetry axis are similar to the separation between the
radio knots behind the main hotspot (assuming that the latter are due
to internal recollimation shocks). Estimates of the Str\"omgren column
show that the dense clouds produced in this simulation could be
partially neutral. They might represent cores for jet induced star
formation in the NLR of Seyfert galaxies.

\section{Acknowledgements}
We thank F. Kahn, J.E. Dyson, J. Meaburn, and D.J. Axon for useful
discussions. We also thank the referee S.A.E.G. Falle for useful 
suggestions. WS, JLG and RJRW acknowledge the receipt of a PPARC
associateship. JLG also gratefully acknowledges a ``Contrato de
Reincorporaci\'on'' by the Spanish Ministry of Education.


\end{document}